\documentclass[final,5p,times,twocolumn,sort&compress,number]{elsarticle}
\usepackage{hyperref}

\usepackage{amsmath}
\usepackage{amssymb}
\usepackage{amstext}
\usepackage{cases}
\usepackage{mathrsfs}
\usepackage{mathtools}
\usepackage{pifont}
\usepackage{ragged2e}
\usepackage{setspace}

\newcommand{\be}{\begin{equation}}
\newcommand{\ee}{\end{equation}}
\newcommand{\bea}{\begin{eqnarray}}
\newcommand{\eea}{\end{eqnarray}}

\newcommand{\gf}[1]{\Gamma\left(#1\right)}
\newcommand{\abs}[1]{\left|#1\right|}
\newcommand{\cc}[1]{\left(#1\right)^{*}_{}}
\newcommand{\isdef}{\mathrel{\mathop:}=}

\newcommand{\tno}[1]{{\theta^{}_{#1}}}

\newcommand{\lble}[1]{\label{e:#1}}
\newcommand{\eqr}[1]{\eqref{e:#1}}

\begin{document}
%\linespread{1.5}

\begin{frontmatter}

\title{The global rotating scalar field vacuum on anti-de Sitter space-time}
\author{Carl Kent}\ead{c.kent@sheffield.ac.uk}
\author{Elizabeth Winstanley}\ead{e.winstanley@sheffield.ac.uk}
\address{Consortium for Fundamental Physics, School of Mathematics and Statistics, University of Sheffield,\\ Hicks Building, Hounsfield Road, Sheffield, S3 7RH, United Kingdom}

\begin{abstract}
We consider the definition of the global vacuum state of a quantum scalar field on $n$-dimensional anti-de Sitter space-time as seen by an observer rotating
about the polar axis.
Since positive (or negative) frequency scalar field modes must have positive (or negative) Klein-Gordon norm respectively,
we find that the only sensible choice of positive frequency corresponds to positive frequency as seen by a static observer.
This means that the global rotating vacuum is identical to the global nonrotating vacuum.
For $n\ge 4$, if the angular velocity of the rotating observer is smaller than the inverse of the anti-de Sitter radius of curvature, then modes with positive Klein-Gordon norm also have positive frequency as seen by the rotating observer.
We comment on the implications of this result for the construction of global rotating thermal states.
\end{abstract}

\end{frontmatter}

\section{Introduction}
\label{int}

In the canonical quantization approach to quantum field theory (QFT), states of the quantum field containing particles are built up from the vacuum state using particle creation operators.
The definition of particle states therefore relies on the definition of a vacuum state (a state with no particles).
On a curved space-time, in general there is no unique definition of vacuum state, although there may be one or more natural, physically motivated, choices of vacuum.

This can be understood by considering the expansion of a free quantum field as a complete orthonormal set of field modes.
Each mode can be classified as either a positive (or negative) frequency mode,
whose expansion coefficient is an annihilation (or a creation) operator respectively.
Since the vacuum state is defined as the state annihilated by all the annihilation operators,
its definition therefore depends on the split into positive and negative frequency field modes.
In the case of a scalar field, the choice of split into positive and negative frequency field modes is constrained by the fact that positive (or negative) frequency modes must have positive (or negative) Klein-Gordon norm respectively.

The consequences of this constraint on the definition of a vacuum state for a quantum scalar field can be illustrated by the simple toy model of Minkowski space as seen by an observer rotating about the polar axis.
In this case the rotating vacuum is identical to the Minkowski vacuum \cite{L&P}.
The constraint on the definition of the vacuum state also has an impact on the definition of states containing particles.
In particular, rotating thermal states for scalar fields in Minkowski space are ill-defined everywhere unless the system is enclosed inside a time-like boundary sufficiently close to the axis of rotation \cite{L&P,D&O}.

Motivated by the results of \cite{L&P,D&O}, in this letter we consider a quantum scalar field on $n$-dimensional global anti-de Sitter space-time ($adS\!^{}_{n}$). We study the constraints on the definition of an appropriate vacuum state as seen by an observer rigidly rotating about the polar axis.
We find that, as in Minkowski space, the global rotating vacuum is identical to the  global nonrotating vacuum.
However, if the angular velocity of the rotating observer is sufficiently small and $n\ge 4$, this global vacuum contains only positive frequency particles as seen by the rotating observer.

\section{Anti-de Sitter space-time in rotating co-ordinates}
\label{ads}

A convenient dimensionless coordinate system for global $adS\!^{}_{n}$ is the set of hyperspherical coordinates\footnote{Throughout this paper we use units in which $c=G=\hbar=1$.},
\be \lble{cs}
\begin{array}{rcll}
-\pi<&\tau&\leq\pi,&\quad\tau=-\pi\text{ and }\tau=\pi\text{ identified,}\\
0\leq&\rho&<\tfrac{\pi}{2},&\\
0\leq&\tno{j}&\leq\pi,&\quad j=1,2,\ldots,n-3,\\
0\leq&\varphi&<2\pi,&
\end{array}
\ee
parametrizing the temporal, radial, polar and azimuthal directions respectively. The coordinate system \eqr{cs} covers $adS\!^{}_{n}$, excluding polar singularities.  In terms of the coordinates \eqr{cs}, the metric on $adS\!^{}_{n}$ takes the form
\be
\lble{nonrot}
ds^{2}_{}=a^{2}_{}\left(\sec\rho\right)^{2}_{}\left[-d\tau^{2}+d\rho^{2}_{}+\left(\sin\rho\right)^{2}_{} d\Sigma _{n-2}^{2} \right] ,
\ee
where $a$ is the radius of curvature of $adS\!^{}_{n}$ and $d\Sigma_{n-2}^{2}$ is the metric on the $(n-2)$-sphere.
Since $\tau = - \pi$ and $\tau = \pi $ are identifed, $adS\!^{}_{n}$ admits closed timelike curves.
To remedy this, we work on  the covering space $C\!adS\!^{}_{n}$ where $-\infty < \tau < +\infty $.

We now consider global $CadS\!^{}_{n}$ as seen by an observer rotating with a constant angular velocity $\Omega $ about the polar axis.
The line-element for the rotating space-time is found from \eqr{nonrot} by the change of co-ordinates
\be
\tau\mapsto \tilde{\tau},\qquad
\varphi\mapsto \tilde{\varphi}\isdef\varphi-\Omega a\tau
\lble{sqphi}
\ee
and takes the form
\begin{align}
\lble{rot}
ds^{2} = & \; a^{2}_{} \left( \sec \rho \right) ^{2}_{}
\left[-\left( 1 - \Omega ^{2}a^{2}{\mathcal {D}}^{2} \left( \frac {\sin \rho }{\rho }\right) ^{2} \right) d{\tilde {\tau }}^{2}
\right. 
\nonumber
\\
 & \;
 \left.
+2\Omega a {\mathcal {D}}^{2} \left( \frac {\sin \rho }{\rho } \right) ^{2} d{\tilde {\tau }} \, d{\tilde {\varphi }}
+d\rho^{2}_{}+\left(\sin\rho\right)^{2}_{} d{\tilde {\Sigma }}_{n-2}^{2} \right] ,
\end{align}
where $d{\tilde {\Sigma }}_{n-2}^{2}$ is the metric on the $(n-2)$-sphere with $d\varphi $ replaced by $d{\tilde {\varphi }}$
and ${\mathcal {D}}$ is the distance from the rotation axis:
\be
\lble{D}
{\mathcal {D}}= \rho\sin\tno{1}\sin\tno{2}\ldots\sin\tno{n-3}.
\ee
The speed of a rotating observer who has angular speed $\Omega $ about the polar axis increases as the distance ${\mathcal {D}}$  from the polar axis increases, and becomes equal to the speed of light when $g_{{\tilde {\tau}}{\tilde {\tau }}}=0$.
This surface is known as the speed-of-light surface (SOL).
At the SOL we have $\Omega ^{2}a^{2}{\mathcal {D}}^{2} \rho ^{-2} \sin ^{2} \rho =1$, and so, from \eqr{D}, if $\Omega a<1$ there is no SOL; if $\Omega a =1$ the SOL is on the equator at the boundary of the space-time and if $\Omega a >1$ the SOL moves closer to the rotation axis as $\Omega $ increases.
Sketches of the SOL can be found in Fig.~1 of \cite{Ambrus:2014fka}.

\section{Scalar field on global anti-de Sitter space-time}
\label{scalar}

The equation of motion for a real massive free scalar field $\Phi(x)$ coupled to $g^{}_{\mu\nu}$, the metric tensor of global $CadS\!^{}_{n}$, is
\be \lble{hkge}
\left(\Box-M^{2}_{}-\xi\mathcal{R}\right)\Phi=0,
\ee
where
\be
\Box\isdef g^{\mu\nu}_{}\nabla^{}_{\mu}\nabla^{}_{\nu}
\ee
is the $n$-dimensional curved-space Laplacian, $M$ is the mass of the field quanta, and the constant $\xi$ is the coupling between $\Phi$ and $\mathcal{R}$, the Ricci scalar curvature.

Solving the Klein-Gordon equation \eqr{hkge} on the nonrotating  global $CadS\!^{}_{n}$ metric \eqr{nonrot}, the mode solutions take the form \cite{Cota1}
\be
\lble{nfm}
\Phi^{}_{r\ell}=N^{}_{r\ell}e^{-i\omega\tau}_{}R(\rho)Y^{}_{\ell}(\theta,\varphi),
\ee
where $N^{}_{r\ell }$ is a normalization constant.
The hyperspherical harmonics $Y^{}_{\ell }(\theta , \varphi )$ are normalized eigenfunctions of the Laplacian on the $(n-2)$-sphere, whose eigenvalues depend
on the angular quantum number $\ell $, which takes the values $\ell = 0,1,2,\ldots $.
For each $\ell $ there are ${\mathcal {M}}^{}_{\ell }$ eigenfunctions, where the multiplicity ${\mathcal {M}}^{}_{\ell }$ is
\cite{Erd2,Muller}
\be \lble{mol}
{\mathcal {M}}^{}_{\ell }=(2\ell+n-3)\frac{(\ell+n-4)!}{\ell!(n-3)!}.
\ee
It will be convenient for our later analysis to separate out the dependence of $Y^{}_{\ell }(\theta , \varphi)$ on the azimuthal angle $\varphi $, so we write
\be \lble{hh}
Y^{}_{\ell}(\theta,\varphi)=e^{\pm im\varphi}_{}\Theta^{}_{\ell m}(\theta),
\ee
where $m\ge 0$ is the azimuthal quantum number and
\be
\theta\isdef\left(\tno{1},\tno{2},\ldots\tno{n-3}\right).
\ee
The function $\Theta ^{}_{\ell m}(\theta )$ also depends on additional quantum numbers associated with the angles $\theta _{2},\ldots , \theta _{n-3}$,
which we denote $m_{1},\ldots ,m_{n-4}$. For compactness of notation, we do not explicitly write out this dependence.
These additional quantum numbers satisfy the inequalities \cite{Erd2,Muller}
\be
\lble{aqn}
\ell \geq m^{}_{1}\geq\ldots\geq m^{}_{n-4}\geq m \ge 0.
\ee

Although $CadS\!^{}_{n}$ does not have any closed time-like curves, it is not a globally hyperbolic space-time because of the time-like boundary at $\rho =\tfrac {\pi }{2}$.
In order to have a well-defined QFT in the next section, appropriate boundary conditions have to be placed on the scalar field $\Phi $ \cite{AS&I}.
We consider regular modes \cite{B&F} which satisfy reflective boundary conditions $\Phi =0$ on $\rho = \tfrac {\pi }{2}$.
These modes exist provided
\be
\lble{k}
k = {\sqrt {M_{}^{2}a_{}^{2}+\xi\mathcal{R}a^{2}_{}+\frac{(n-1)_{}^{2}}{4} }} + \frac {n-1}{2}
\ee
is real.
With this assumption, the radial function in \eqr{nfm} takes the form
\be
\lble{rm}
R(\rho)\isdef(\sin\rho)^{\ell}_{}(\cos\rho)^{k}_{}P^{\,\left(\ell+\frac{n-3}{2},\,k-\frac{n-1}{2}\right)}_{r}\left(\cos(2\rho)\right),
\ee
where $P^{\,\left(\ell +\frac{n-3}{2},\,k-\frac{n-1}{2}\right)}_{r}$ is a Jacobi polynomial of degree $r$ and we have introduced the radial quantum
number $r=0,1,\ldots $.

The modes \eqr{nfm} are normalized according to the Klein-Gordon inner product:
\be
\lble{KGinner}
\langle\Phi^{}_{r\ell},\Phi^{}_{r'\ell'}\rangle^{}_{\text{KG}}=-\int^{}_{H}d^{n-1}_{}\boldsymbol{x}\sqrt{g}g^{\tau\tau}_{}\cc{\Phi^{}_{r\ell}} \overset{\leftrightarrow}{\partial^{}_{\tau}}\Phi^{}_{r'\ell'},
\ee
evaluated on some space-like hypersurface of simultaneity $H$, with
\be
A\overset{\leftrightarrow}{\partial}^{}_{\mu }B\isdef A\partial^{}_{\mu}B-(\partial^{}_{\mu}A)B,
\ee
and
\be
g\isdef\abs{\det g^{}_{\mu\nu}}.
\ee
The normalization constant $N^{}_{r\ell }$ is then found to be \cite{Cota1}
\be \lble{N}
N^{}_{r\ell}=a_{}^{\frac{2-n}{2}}\,\sqrt{\frac{r!\gf{r+\ell+k}}{\gf{r+\ell+\frac{n-1}{2}}\gf{r+k-\frac{n-3}{2}}}}.
\ee

\section{Defining a global nonrotating vacuum}
\label{vac}

As outlined in the Introduction, the first step in defining a global vacuum state is to split the field modes into positive and negative frequency.
We start by considering the nonrotating modes \eqr{nfm}.
These modes have frequency $\omega $ as seen by a static observer in global $CadS\!^{}_{n}$.
Computing their Klein-Gordon inner product \eqr{KGinner}, we find
\be
\langle\Phi^{}_{r\ell},\Phi^{}_{r'\ell'}\rangle^{}_{\text{KG}}=\frac {\omega }{\left| \omega  \right|} \delta^{}_{rr'}\delta^{}_{\ell\ell'}.
\lble{KGnorm}
\ee
Therefore modes with positive $\omega $ have positive Klein-Gordon norm, while those with negative $\omega $ have negative norm.
We therefore take $\omega >0$ as our definition of positive frequency.
With this assumption, for $n\ge 4$ the frequency $\omega $ is given in terms of the radial and angular quantum numbers \cite{Cota1}:
\be \lble{o}
\omega=k+\ell+2r ,
\ee
which is manifestly positive as $k$ \eqr{k}, $\ell, r$ are all positive.
We discuss the $n=3$ case in the next section.

The quantum scalar field is expanded in terms of these modes as
\be
\Phi = \sum _{r=0}^{\infty }\sum _{\ell =0}^{\infty} \sum _{m,m_{1},\ldots, m_{n-4}}
\left[
b^{}_{r\ell }\Phi ^{}_{r\ell } + b^{\dagger }_{r\ell }{\cc {\Phi ^{}_{r\ell }}}
\right] ,
\lble{exp}
\ee
where $m_{1},\ldots ,m_{n-4}$ are additional quantum numbers arising in the spherical harmonics \eqr{hh}.
We have suppressed the dependence of $\Phi _{r\ell }$ and $b^{}_{r\ell }$ on these additional quantum numbers just to keep the notation compact.
Quantizing the field, the coefficients $b^{}_{r\ell }$ and $b^{\dagger }_{r\ell }$ are promoted to operators satisfying the usual
commutation relations:
\be
[ b^{}_{r\ell }, b^{\dagger }_{r'\ell'} ] = \delta _{rr'}\delta _{\ell \ell'} \delta \left(m, m'\right) ,
\quad
[ b^{}_{r\ell }, b{}_{r'\ell'}] = 0 = [ b^{\dagger }_{r\ell }, b^{\dagger }_{r'\ell '} ],
\ee
where we have introduced the notation
\be
\delta \left (m,m' \right) = \delta _{mm'}\delta _{m_{1},m_{1}'} \ldots \delta _{m_{n-4},m_{n-4}'}.
\ee
The global nonrotating vacuum state $\left| 0 \right\rangle $ is then defined as that state annihilated by all the $b_{r\ell }$ operators:
\be
b_{r\ell }\left| 0 \right\rangle = 0.
\lble{vac}
\ee
This vacuum state has been studied in detail in \cite{KW-pI}, where the expectation values of the renormalized quadratic field fluctuations and
stress-energy tensor are computed.

\section{Defining a global rotating vacuum}
\label{rot}

Now we turn to the definition of a global rotating vacuum state. Scalar field modes on the rotating global $CadS\!^{}_{n}$ metric \eqr{rot} are easily found from those on the nonrotating metric \eqr{nonrot} by making the coordinate transformation \eqr{sqphi} in the modes \eqr{nfm}, yielding
\be \lble{rfm}
\tilde{\Phi}^{}_{r\ell}(x)=N_{r\ell}e^{- i\tilde{\omega}\tilde{\tau}}_{}R(\rho)e^{im\tilde{\varphi}}_{}\Theta^{}_{\ell m}(\theta),
\ee
where
\be
\lble{sqo}
\tilde{\omega}\isdef\omega-\Omega a m.
\ee
An observer rotating about the polar axis with angular velocity $\Omega $ measures the frequency of the modes \eqr{nfm} to be ${\tilde {\omega }}$ \eqr{sqo}.
In this case, it is natural to consider the modes in the alternative form \eqr{rfm}.
However, our choice of positive frequency is restricted by the fact that positive frequency modes must have positive Klein-Gordon norm.
From \eqr{KGnorm}, the only possible choice of positive frequency is $\omega >0$.
We therefore expand the field as in the nonrotating case \eqr{exp}, and end up with the global nonrotating vacuum $\left| 0 \right\rangle $ \eqr{vac}.

In Minkowski space, the set of modes with positive Klein-Gordon norm always contains some modes which have negative frequency as seen by an observer rotating
about the polar axis.
This has serious consequences for the construction of states containing particles, and, in particular, rotating thermal states.
The rotating observer measures energy ${\tilde {\omega }}$ for the field modes, and so the natural definition of a rotating thermal state will have
energy ${\tilde {\omega }}$ in the Planck factor \cite{Vilenkin:1980zv}.
However, this definition leads to rotating thermal states for a quantum scalar field being ill-defined everywhere in Minkowski space-time \cite{Vilenkin:1980zv,D&O,A&Wro}.
The only solution to this problem is to enclose the system inside a time-like boundary which is sufficiently close to the axis of rotation \cite{Vilenkin:1980zv,D&O}.
The inclusion of the boundary solves the problem by ensuring that modes with positive Klein-Gordon norm also have positive frequency as seen by the
rotating observer.

Given that $CadS\!^{}_{n}$ has a time-like boundary at $\rho = \tfrac {\pi }{2}$, the question arises as to whether modes with positive Klein-Gordon norm on
$CadS\!^{}_{n}$ can have negative frequency as seen by an observer rotating about the polar axis with angular velocity $\Omega $.
In other words, are there field modes \eqr{rfm} which have $\omega >0$ but ${\tilde {\omega }}<0$?
For $n\ge 4$, we note that $\omega >0$ is given in terms of the quantum numbers $r$ and $\ell $ \eqr{o}, and from this the inequalities \eqr{aqn}
imply that, for $\omega >0$, we have
\be
\omega \ge k + 2r + m > m ,
\ee
since $k>0$ \eqr{k}.
Hence, from \eqr{sqo}
\be
{\tilde {\omega }} = \omega - \Omega a m > m \left( 1- \Omega a \right) .
\ee
Therefore, if $\Omega a <1$, it will be the case that modes with positive Klein-Gordon norm also have positive frequency as seen by the rotating observer.
If $\Omega a <1$, then, from the discussion in Sec.~\ref{ads}, the rotating space-time does not have a SOL.

Our results on global $CadS\!^{}_{n}$ for $n\ge 4$ therefore agree with those in rotating Minkowski space \cite{Vilenkin:1980zv,D&O}: if there is no SOL, then modes with
positive Klein-Gordon norm have positive frequency as seen by the rotating observer.
In Minkowski space showing this result depends on the properties of the zeros of Bessel functions \cite{D&O}, whereas in $CadS\!^{}_{n}$ it comes from the relationship between the mode frequency and the quantum numbers, and the inequalities \eqr{aqn} satisfied by the angular quantum numbers.

The situation on global $CadS\!^{}_{3}$ is slightly different.  In order that positive frequency modes have positive
Klein-Gordon norm, we must still have $\omega >0$ as the definition of positive frequency.
This means that the only choice of global vacuum state remains the global nonrotating vacuum.
However, for $n=3$ the frequency $\omega $ depends on the azimuthal quantum number $m\ge 0$ as follows \cite{Parikh1}:
\be
\omega = k + 2r \pm m,
\lble{o3}
\ee
so that
\be
{\tilde {\omega }} = \omega - ma\Omega = k + 2r - m \left( a\Omega \mp 1 \right) .
\lble{o3t}
\ee
Therefore there exist, for sufficiently large $m$, counter-rotating modes (corresponding to the lower signs in (\ref{e:o3}, \ref{e:o3t})) which have
$\omega >0$ but ${\tilde {\omega }}<0$ \cite{Parikh1}.
Such modes have negative frequency as seen by the rotating observer and, as discussed above, are anticipated to render
rotating thermal states ill-defined.
In \cite{Parikh1} an alternative vacuum state is defined when $n=3$ for a rotating anti-de Sitter space-time which has a cylindrical region near the axis of rotation removed.
There is also a family of alternative vacua on rotating Rindler-$adS\!^{}_{3}$ space-time \cite{Parikh1}.
Rotating Rindler-$adS\!^{}_{3}$ possesses an event horizon and corresponds to a portion of the global
$adS\!^{}_{3}$ space-time in the same way that the usual Rindler space-time is only a part of global Minkowski space-time.
In this paper we are considering the entire global $CadS\!^{}_{n}$ space-time and the alternative vacuum states from \cite{Parikh1} cannot be defined in this case.

\section{Conclusions}
\label{conc}

In this paper we have studied a quantum scalar field on global $CadS\!^{}_{n}$ as seen by an observer rotating about the polar axis with angular velocity $\Omega $.
We found that the requirement that positive frequency modes must have positive Klein-Gordon norm (to ensure that the particle annihilation and creation operators satisfy the correct commutation relations) restricts our choice of vacuum state, so that the only possibility is the global nonrotating vacuum.
If $n\ge 4$ and the angular velocity satisfies the inequality $\Omega a <1$ (where $a$ is the radius of curvature of $adS\!^{}_{n}$), then scalar field modes with positive Klein-Gordon norm also have positive frequency as seen by the rotating observer.
In this case the global nonrotating vacuum is the natural state to use for constructing states which contain particles as seen by the rotating observer.
It is of note that if $\Omega a<1$ then the rotating $CadS\!^{}_{n}$ space-time does not have a speed-of-light surface (SOL).

Our results are in accordance with previous work on quantum scalar fields on rotating Minkowski space-time, in particular (i) the global rotating vacuum is identical to the global nonrotating vacuum, and (ii) if the space-time does not have a SOL (in Minkowski space this is achieved by enclosing the system in a boundary sufficiently close to the axis of rotation) then modes with positive Klein-Gordon norm have positive frequency as seen by the rotating observer.

In Minkowski space, if the boundary is inside the SOL, then it is possible to define rotating thermal states for a quantum scalar field, but such states are ill-defined everywhere if the boundary is either outside the SOL or absent \cite{D&O}.
We expect that similar results will be true in $CadS\!^{}_{n}$ with $n\ge 4$: that if $\Omega a <1$ then rotating thermal states are well-defined for a quantum scalar field, but they are not if $\Omega a \ge 1$.
We will investigate this in detail in a future publication \cite{KWaip}.

\section*{Acknowledgments}

The work of C.K. is supported by EPSRC UK, while that of E.W. is supported by the Lancaster-Manchester-Sheffield Consortium for Fundamental Physics under STFC grant ST/L000520/1.

%\section*{References}


\begin{thebibliography}{10}
\expandafter\ifx\csname url\endcsname\relax
  \def\url#1{\texttt{#1}}\fi
\expandafter\ifx\csname urlprefix\endcsname\relax\def\urlprefix{URL }\fi
\expandafter\ifx\csname href\endcsname\relax
  \def\href#1#2{#2} \def\path#1{#1}\fi

\bibitem{L&P}
J.~R. {Letaw}, J.~D. {Pfautsch}, Quantized scalar field in the stationary
  coordinate systems of flat spacetime, Phys. Rev. D 24 (1981) 1491--1498.

\bibitem{D&O}
G.~{Duffy}, A.~C. {Ottewill}, Rotating quantum thermal distribution, Phys. Rev.
  D 67 (2003) 044002.
%

\bibitem{Ambrus:2014fka}
V.~E. Ambrus, E.~Winstanley, {Dirac fermions on an anti-de Sitter background \,
  }\href {http://arxiv.org/abs/1405.2215} {\path{arXiv:1405.2215}}.

\bibitem{Cota1}
I.~I. {Cot\u{a}escu}, Remarks on the quantum modes of the scalar field on
  {$AdS_{d+1}$} spacetime, Phys. Rev. D 60 (1999) 107504.

\bibitem{Erd2}
A.~{Erd\'{e}lyi} (Ed.), Higher transcendental functions, Vol.~2, McGraw-Hill,
  New York, 1953.

\bibitem{Muller}
C.~{M\"{u}ller}, Spherical harmonics, Lecture notes in mathematics,
  Springer-Verlag, Berlin/ Heidelberg, 1966.

\bibitem{AS&I}
S.~J. {Avis}, C.~J. {Isham}, D.~{Storey}, Quantum field theory in anti-de
  {S}itter space-time, Phys. Rev. D 18 (1978) 3565--3576.

\bibitem{B&F}
P.~{Breitenlohner}, D.~Z. {Freedman}, Stability in gauge extended supergravity,
  Ann. Phys. (N. Y.) 144 (1982) 249--281.

\bibitem{KW-pI}
C.~{Kent}, E.~{Winstanley}, Hadamard renormalized scalar field theory on
  anti-de {S}itter space-time \,\href {http://arxiv.org/abs/1408.6738}
  {\path{arXiv:1408.6738}}.

\bibitem{Vilenkin:1980zv}
A.~Vilenkin, {Quantum field theory at finite temperature in a rotating system},
  Phys. Rev. D 21 (1980) 2260--2269.

\bibitem{A&Wro}
V.~E. Ambrus, E.~Winstanley, {Rotating quantum states}, Phys. Lett. B 734
  (2014) 296--301.

\bibitem{Parikh1}
M.~Parikh, P.~Samantray, E.~Verlinde, {Rotating Rindler-AdS space},
  Phys. Rev. D 86 (2012) 024005.

\bibitem{KWaip}
C.~{Kent}, E.~{Winstanley}, \,article in preparation.

\end{thebibliography}
\end{document}